\documentclass[letterpaper]{article} 
\usepackage{aaai24}  
\usepackage{times}  
\usepackage{helvet}  
\usepackage{courier}  
\usepackage[hyphens]{url}  
\usepackage{graphicx} 
\usepackage{natbib}  
\usepackage{caption} 
\usepackage{booktabs}
\usepackage{algorithm}
\usepackage{algorithmic}

\usepackage{newfloat}
\usepackage{listings}
\DeclareCaptionStyle{ruled}{labelfont=normalfont,labelsep=colon,strut=off} 
\floatstyle{ruled}
\newfloat{listing}{tb}{lst}{}
\floatname{listing}{Listing}
\title{Time Series Analysis of Key Societal Events as Reflected in Complex Social Media Data Streams}
\author {
    Andy Skumanich\textsuperscript{\rm 1},
    Han Kyul Kim\textsuperscript{\rm 2}
    }

\affiliations {
    \textsuperscript{\rm 1}Innov8ai Inc.
    
    413 Hershner Dr.
    
    Los Gatos, California, 95032-4002 USA\\
    
    \textsuperscript{\rm 2}University of Southern California \\
    ASkuman@Innov8ai.com, HanKyulK@USC.edu
}

\usepackage{bibentry}

\begin{document}

\maketitle

\begin{abstract}
Social media platforms hold valuable insights, yet extracting essential information can be challenging. Traditional top-down approaches often struggle to capture critical signals in rapidly changing events. As global events evolve swiftly, social media narratives, including instances of disinformation, become significant sources of insights. To address the need for an inductive strategy, we explore a \textit{niche} social media platform GAB and an established messaging service Telegram, to develop methodologies applicable on a broader scale. This study investigates narrative evolution on these platforms using quantitative corpus-based discourse analysis techniques. Our approach is a novel mode to study multiple social media domains to distil key information which may be obscured otherwise, allowing for useful and actionable insights. The paper details the technical and methodological aspects of gathering and preprocessing GAB and Telegram data for a \textit{keyness} (Log Ratio) metric analysis, identifying crucial nouns and verbs for deeper exploration. Empirically, this approach is applied to a case study of a well defined event that had global impact: the 2023 Wagner mutiny.  The main findings are: (1) the time line can be deconstructed to provide useful data features allowing for improved interpretation; (2) a methodology is applied which provides a basis for generalization.  The key contribution is an approach, that in some cases, provides the ability to capture the dynamic narrative shifts over time with elevated confidence. The approach can augment near-real-time assessment of key social movements, allowing for informed governance choices. This research is important because it lays out a useful methodology for time series relevant info-culling, which can enable proactive modes for positive social engagement.
\end{abstract}

\section{Introduction}

Social networks have had a profound impact on the way we communicate, share information, and interact with each other. They have become a central part of modern society, enabling people to connect with each other regardless of their location and participate in online communities \cite{hromic10.1007/s13278-019-0576-8}. However, the rise of social networks has also led to several challenges, including the spread of disinformation \cite{10.1145/3543873.3587348}, cyberbullying \cite{GIUMETTI2022101314}, and the propagation of hate speech and extremist ideologies \cite{10.1145/3583067}, along with in general what we call \textit{mal-info} \cite{10.1080/01972243.2022.2139031}. \textit{Mal-info} strategies encompass the use of social media to spread rumors, mobilize protests, and disseminate false narratives. For instance, in the broader geopolitical context of the Wagner scenario, a Telegram post by an alleged Wagner operative claimed France was involved in the \textit{mass removal of children} for sinister purposes. These instances of disinformation could be detected and addressed.\footnote{\url{https://apnews.com/article/niger-coup-jihadis-west-africa-9032a0e1161551ffcfde4b785f6cf74a}}. 


The time series analysis of social media has garnered attention given the challenges of extracting meaningful signal-to-noise.  Several recent papers have highlighted these challenges.\footnote{Wang, et. al. https://doi.org/10.1016/j.neucom.2021.04.020}
As well as others where they investigate the problem of mining timelines of entities in social media.\footnote{Hills, et. al. http://dx.doi.org/10.18653/v1/2023.eacl-main.274
}

Amid the challenges of information overload on social media, including platforms like GAB or Telegram, there is a growing interest in developing inductive, machine-guided methods to mine narratives from vast amounts of posts. Early examples of computational narrative analysis can be traced back to work on artificial intelligence and story structures. Recent research has further explored the underlying dynamics of narratives by representing them as evolving networks of relations between key actors identified through Named Entity Recognition (NER)\cite{10.48550/arxiv.cs/0205028}.

Co-occurrence networks \cite{edwards2018comparing} and empirically-informed approaches have shed light on the structural connections that allow online conspiracy theories and narratives to be discerned from seemingly unrelated concepts and information.

This paper aims to assist the study of online narratives, emphasizing the significance of bottom-up methods to identify idiosyncratic and evolving concepts that form these narratives. Bridging cultural-theoretical and computational-linguistic approaches, previous literature has used word embeddings to reveal how platforms like 4chan foster \textit{robust vernacular innovations} \cite{Vernacular2021}. Addressing the need for inductive approaches to narrative evolution on social networks, especially on new ones such as GAB, and messaging platforms, such as Telegram, this paper employs quantitative methods from corpus-based discourse analysis. The technical and methodological aspects include data collection and preprocessing of messages from the two platforms to conduct a \textit{keyness} (Log Ratio) analysis, identifying significant nouns and verbs for further investigation \cite{willaert2022detecting}.
The empirical part of the paper presents a case study using GAB and Telegram datasets\footnote{Both datasets will be made available on open-source platforms.} spanning the Wagner militia mutiny attempt. The study tests the hypothesis that during this event, GAB posts that previously spread disinformation narratives on subjects like the War in Ukraine or US domestic issues, would embrace misdirecting narratives about the mutiny, and be likely to reflect the Russian discourses instead of the Ukrainian one.
The paper concludes with a broader reflection on the potential and future prospects of this overview approach to narratives in conjunction with established interpretative practices.

\section{Data Collection}

\subsection{The social network GAB}


Several new social networks have developed recently in response to reactions about perceived \textit{censorship} and moderation on already established social media platforms. Some users feel that their \textit{freedom of speech} is being limited on platforms such as Twitter and Facebook and that their content is being targeted or removed.  This is a broader discussion beyond the scope of this paper as it is part of the determination of what constitutes \textit{free speech}. As shown by \cite{stocking2022role}, more and more Americans are using these platforms for news ($6\%$ in 2022). Using tools like Similarweb\footnote{https://www.similarweb.com/}, we can extrapolate that this number must have doubled in 2023.

These new niche platforms often have little to no moderation and highly lenient content policies.  Although this allows for a wider range of opinions and viewpoints to be shared, some of which can indeed be mal-info. However, they can be legitimately criticized for allowing hate speech, harassment \cite{abarna2022identification}, and disinformation to spread unchecked. In this myriad of new social networks, the \textit{GAB} (or Gab) platform is one of particular interest as a newly frequently used channel in the unfettered dissemination mode.   

GAB\footnote{https://gab.com} is a social networking platform launched in 2016 and bills itself as an unfettered speech (so-called \textit{free speech}) alternative to mainstream social media sites. It was created in response to the perceived censorship of conservative views on traditional social media sites. GAB allows users to post messages called \textit{gabs}, share photos, and interact with other users. It has been observed as being a platform for hate speech and far-right extremism.

\subsection{Telegram}

Telegram is a cloud-based instant messaging app and social media platform. It was developed by Pavel Durov and his brother Nikolai Durov and was first launched in 2013. Telegram allows users to send text messages, voice messages, multimedia files, and make voice and video calls. One of its distinguishing features is its focus on security and privacy, offering end-to-end encryption for messages and a self-destruct timer for messages. Telegram also supports group chats, channels, and the ability to create public or private communities for sharing content and interacting with others. It is available on various platforms including smartphones, tablets, and desktop computers. Telegram has now more than 700 million global monthly active users and it is increasingly seen as a source of information \cite{10.1142/S0219649221500246}.

\subsection{Data Collection}

The focus of this paper is on the analysis of narrative time series evolution in posts on GAB and Telegram for extracting meaningful insights.  Examined in particular is the Wagner mutiny in Russia that happened on the 24th of June 2023. These events are of particular note because of both the specific events as well as the broader geo-political implications. Importantly from a time series data analysis perspective, the event represents a very clearly defined occurrence similar to a mathematical \textit{delta function} which is a singular pulse in time.   Given this delta function aspect, the responses can be better observed in the time series developments.  It is essential to note the exploratory nature of this work and that although it is based on this single event, it allows for greater clarity in the analysis, and provides a starting point for more expansive studies.  Our follow-on studies expand on this initial investigation.

To retrieve the necessary time series data, as no API is available for GAB, a tailor-made scraper based on Python’s Selenium library \cite{10.5555/2655462} was used to automate and scale up this process. 
We then retrieved all posts containing $\#$Wagner, $\#$Russia and $\#$Ukraine from the 22nd to the 26th of June 2023. Table \ref{tab:distuposts} shows the number of posts by hashtag. In total, we retrieved 2061 unique posts. We can see that only a minority of posts contain the $\#$Wagner, with users concentrating more on the terms $\#$Russia and $\#$Ukraine during this period. 

\begin{table}[htbp]
\centering
\caption{Number of posts by hashtag from the 22nd to the 26th of June 2023.}
  \label{tab:distuposts}
\begin{tabular}{rrrrrrrrrrrrrrr}
\toprule
Hashtag& Nb. of posts\\ 
  \midrule
$\#$Wagner&340\\ 
$\#$Russia&923\\ 
$\#$Ukraine&798\\ 
 \midrule
Total&2061\\ 
\bottomrule
\end{tabular}

\end{table}

\begin{figure}[!t]
\centering
\includegraphics[width=1\linewidth]{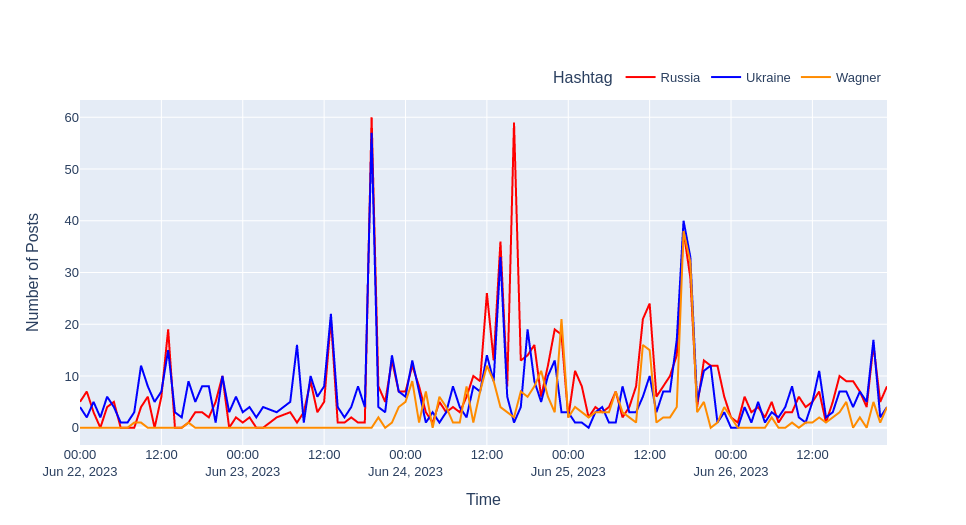}
\caption{Distribution of posts by hashtags over time (UTC time) for GAB.}
\label{timelineHasthags}
\end{figure}

Figure \ref{timelineHasthags} shows the number of posts over time. It is interesting to note that the $\#$Wagner appeared mainly at the start of the mutiny on 24 June and almost ceased to exist at the end of the day on 26. Just before the end of that day on the 26th, people posted mainly about the agreement between the head of the Wagner Militia and the Russian President. It is also interesting to note that the number of posts containing the hashtags $\#$Russia and $\#$Ukraine during this period is similarly the same except for the time when Wagner militia troops were advancing towards Moscow, around 3 PM UTC on the 24 of June. We suggest a connection between the timing of the peaks and the occurrence of public statements by the key players.  This connection indicates an EU and US set of time zones in terms of responses (note that GAB is US based). The rapid onset and drop-off are notable, showing notable time series behavior in the surge and ebb in GAB activity, the reasons behind which could be further investigated by subsequent multi-domain research including social science.  The extraction of the time series patterns can now allow for cross-domain analysis.  The societal evaluation is dependent on clean data to allow for interpretation, which this case study exemplifies.  This is also a case of data quality instead of data quantity providing the analysis value.  Even with the low post count a good singnal-to-noise is extracted.

We then applied a classical preprocessing step in which we removed punctuation, number and stopwords \cite{barbaro:tel-03708173}. 
After preprocessing, it was found that the retrieved messages all contained text.  Automated language detection, using the \textit{langdetect} library \cite{shuyo2010language}, revealed that the corpus was multilingual, with English being the most prominent language.  Of the message texts, $86\%$ (1774) were classified as written in English. In what follows, we restrict ourselves to analysing posts in English.

Afterwards, we take a look at the URL that users share as a part of their posts. Of the 1774 posts in English, 1356 posts contained an URL. Within the top 50 domains that were
shared, we find several domains that are known Russia state-sponsored media (www.RT.com), or tagged as questionable (www.rumble.com) and prone to conspiracies and pseudoscience (www.zerohedge.com) by Media Bias / Fact Check
(MBFC)\footnote{https://mediabiasfactcheck.com/}. MBFC is an independently operated website that classifies domains into certain factual and political lean categories, based on specific
criteria that are described further in their per-domain synopsis.

To compare and contrast our GAB results, we retrieved data from Telegram representing the main channels of the belligerents in the war in Ukraine, namely Ukraine and Russia, from the 22nd to the 26th of June 2023. To do so, we used the API provided by Telegram and used the Python package \textit{Telethon}\footnote{https://pypi.org/project/Telethon/.}. We selected two channels for each country in the Russian language.

For Ukraine, we choose Ukraine Now\footnote{\url{https://t.me/u_now}} and Union\footnote{https://t.me/uniannet}, respectively with more than $1725000$ and $850000$ subscribers. Ukraine Now is a source of information entirely based on Telegram while Union is a more traditional online media.

For Russia, we choose Ria Novosti\footnote{\url{https://t.me/rian_ru}} and Union\footnote{https://t.me/kommersant}, respectively with more than $2950000$ and $200000$ subscribers. RIA Novosti is a Russian state-owned domestic news agency and Kommersant is a nationally distributed daily newspaper published in Russia mostly devoted to politics and business with close ties to the Kremlin.

\begin{figure}[!t]
\centering
\includegraphics[width=1\linewidth]{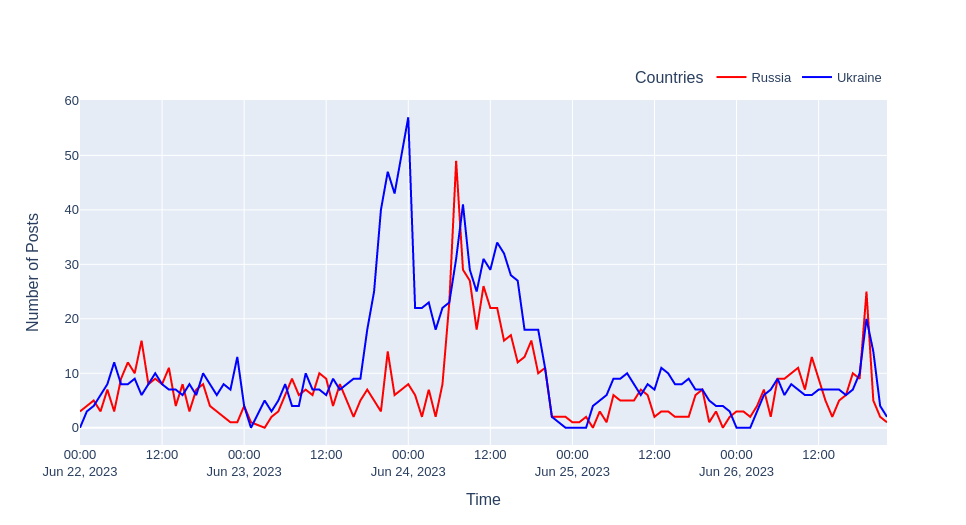}
\caption{Distribution of aggregated posts by countries over time (UTC time) for Telegram.}
\label{timelinePostsTelegram}
\end{figure}

Table \ref{tab:distupoststelegram} shows the number of posts by source and country. In total, we retrieved $1397$ posts for Ukraine and $867$ posts for Russia. It seems that Russian sources were less inclined to talk about the mutiny at first and were perhaps waiting for official directives from the Kremlin. A time delay on the Russia feed is apparent in Figure \ref{timelinePostsTelegram}. Indeed, there is a large gap between the peak of news published by Ukrainian sources and those published by Russian sources, even with the same provoking events. Still, there is a similarity between the peaks of posts present on Gab and on Telegram. 

\begin{table}[htbp]
\centering
\caption{Number of posts by hashtag from the 22nd to the 26th of June 2023.}
  \label{tab:distupoststelegram}
\begin{tabular}{rp{10mm}p{10mm}p{10mm}p{10mm}}
\toprule
& \multicolumn{2}{c}{Ukraine}&\multicolumn{2}{c}{Russia}\\ 
  \midrule
Source&Ukraine Now& Union &Ria Novosti&Komme- rsant\\ 
Nb posts&824&543&591&276\\ 
 \midrule
Total& \multicolumn{2}{c}{1367}&\multicolumn{2}{c}{867}\\ 
\bottomrule
\end{tabular}

\end{table}

\section{Methodology}

To effectively identify signals of narrative evolution in the collected data, this paper employs the method of \textit{keyness} analysis. Drawing inspiration from the fields of corpus linguistics and corpus-based discourse analysis, this approach focuses on identifying 'key' items, such as words or phrases, within a target corpus about a reference corpus. By comparing the frequencies of items in both corpora, \textit{keyness} analysis allows for an exploratory examination of texts, providing insights into their underlying themes and subjects.

One of the primary reasons for choosing \textit{keyness} analysis in this study is its suitability for identifying emerging narrative signals within texts. The \textit{keyness} metric adopted for this paper is the Log Ratio, which is defined as the \textit{binary log of the ratio of relative frequencies} \cite{hardie2014log}. Unlike measures of statistical significance, the Log Ratio provides a measure of the actual observed difference in frequencies between two corpora for a key item. This feature enables the sorting of items based on the size of the frequency difference, facilitating the identification of the top N most key items (emerging from the prior period). As such, a \textit{keyness} analysis can support an exploratory approach to texts that indicates their \textit{aboutness}, i.e. what matters to the posting community\cite{7304f57f57124281ad6df37086a965eb}.

To calculate the Log Ratio for an item in the target corpus (C1) and the reference corpus (C2), the binary logarithm of the ratio of the normalized frequencies of the term in both corpora is taken. For readability purposes, these frequencies are multiplied by a factor of $1,000,000$. The Log Ratio metric allows for a quantitative assessment of the significance of key items in terms of their deviation from the norm.

\begin{figure}[!t]
\centering
\includegraphics[width=1\linewidth]{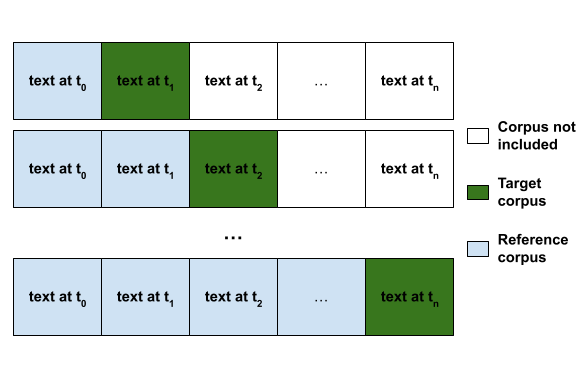}
\caption{Schematic overview of the approach. Data are
grouped by timestamps. For data at each timestamp (the
target corpus), \textit{keyness} scores (Log Ratio) for nouns and
verbs are calculated in relation to data of the reference corpus (previous timestamps). }
\label{temporal_system}
\end{figure}

In the context of narrative detection on GAB and Telegram, the overall approach involves detecting key items from a reference corpus comprising texts grouped by day as illustrated in Figure \ref{temporal_system}. These key items are then examined about the remaining data, enabling further interpretation. This process allows for the identification of the items with the highest \textit{keyness} scores at each timestamp, providing valuable insights into the evolving narratives within the data.

Specifically, the technical pipeline for this analysis consists of the following sequential steps:

\begin{enumerate}
    \item We filter the data to retain only English posts for GAB and Russian posts for Telegram and group posts by timestamps (per day); 

\item In addition to classic preprocessing, we clean posts by
removing hyperlinks and emojis ;
\item We perform part of speech tagging and retain
only nouns and verbs using the library \textit{NLTK} \cite{10.48550/arxiv.cs/0205028};
\item We calculate the frequencies for these items
per timestamp;
\item We calculate the Log Ratio of the target corpus compared to the reference corpus as in Figure \ref{temporal_system};
\item Finally, we rank words by \textit{keyness} score.
\end{enumerate}

In a conceptual sense, this approach generates \textit{keyness} scores for items relative to the preceding data, providing a current perspective on distinct narrative signals for each day's dataset in relation to the entire preceding period. The \textit{keyness} scores for the final timestamp hold particular significance as they unveil key items in relation to all preceding data, shedding light on what is deemed significant at the latest observation moment. It is important to note that this \textit{keyness} analysis does not currently incorporate semantics, except for considering nouns and verbs as key indicators of narratives.

To exemplify this methodology and contribute empirically to the examination of narrative dynamics on GAB and Telegram, the following section focuses on a case study that explores the narratives surrounding the mutiny organized by the Wagner Militia.

\section{Case Studies and Findings}

Recent events, such as the coronavirus pandemic and the war in Ukraine, have sparked interest among researchers, civil society actors, and journalists in understanding the evolving nature of (disinformation) narratives. A comparative analysis of international fact-checks has highlighted notable similarities in terms of style and content between disinformation surrounding these events \cite{5aac817b156b439c920d831d466fbc2a}. Examples include references to Nazism, such as labeling the coronapass as a Nazi \textit{health passport} or advocating for the \textit{denazification} of Ukraine. 
In addition, various recurring conspiracy theories have emerged in these sub-groups, the supposed Covid vaccine effects (micro-chip injections), witch hunts against the former US president, or the supposed cost of aid to Ukraine.
Specifically, a relevant question is - do the same communities that previously spread false narratives about US politic or the coronavirus also engage in disinformation regarding the war in Ukraine? - and, how did they express the events surrounding the Wagner militia mutiny?

According to a recent study conducted by the Institute of Strategic Dialogue \cite{SmirnovaTel2022}, it has been confirmed that this co-aligned narrative phenomenon does occur. The study examined 229 German-language Telegram channels associated with far-right and conspiracy theory communities, covering the period from November 1, 2021, to February 27, 2022. The analysis revealed that terms related to \textit{Russia, Ukraine, the breakaway regions in Eastern Ukraine, and the Russia-Ukraine crisis} from a predefined list of 80 keywords became more coincident and prevalent in the discourse of these communities. The study further examined the actual narratives, particularly those promoting a pro-Russian perspective, through a detailed analysis of articles from the most frequently shared domains within the dataset. The GAB dataset we study in this article differs in the source but concerns the same population. A population close to conspiracy and far-right circles as well as to Russian state argumentation, as we saw earlier with the links shared in the posts. By employing our inductive approach, we hypothesize that we can uncover intricate traces of the actual narratives that emerge. This, in turn, provides us with insights into the underlying dynamics that drive this narrative evolution at a deeper level.

The telegram channels we selected are directly involved in the information war as they present facts from divergent sides of the ground war. It will assist us to understand how a particular group evolves during the Wagner mutiny and if we can find any similarities or an echo in Gab's posts.

\subsection{Gab case study}
To conduct our analysis of narratives, we focus on the four last days of our database, from 23 June to 26 June. Note that the corpus of 22 June serves as the reference corpus for 23 June. This time frame allows for covering key elements of the Wagner militia mutiny, from Prigozhin's allegations of an attack on a Wagner militia base by the regular Russian army\footnote{https://www.bellingcat.com/news/2023/06/23/site-of-alleged-wagner-camp-attack-recently-visited-by-war-blogger/}, on 23 June, to the subsequent outcome of his exile in Belarus and his first speech on 26 June\footnote{https://www.euronews.com/2023/06/26/wagner-leader-prigozhin-breaks-his-silence-issuing-first-audio-statement-since-mutiny} after the mutiny was aborted. 

A first observation on the basis of our \textit{keyness} analysis, is that we can indeed see emerging traces of narratives concerning the attempted mutiny but also more recurrent topics such as the war in Ukraine, US aid policy towards Ukraine and Covid Vaccine. Table \ref{tab:narrativestable} shows the top 10 nouns and verbs (by \textit{keyness} score) retrieved for the last four days of English posts texts in the dataset. It's worth noting how quickly the narratives evolve in response to current events, returning to more traditional hard-right topics such as American politics and the Covid vaccine.

\begin{table}[htbp]
\centering
\caption{Top 10 nouns and verbs with highest \textit{keyness} scores for GAB's message texts in English for the last four days of the dataset (23 June to 26 June).}
  \label{tab:narrativestable}
\begin{tabular}{rrrrrrrrrrrrrrr}
\toprule
2023-06-23& 2023-06-24&2023-06-25&2023-06-26\\ 
  \midrule
Battlefield&Combat&Enters&Weakness\\
Tax&Pmc&Stand&Empire\\
Strike&Soldier&Incident&Class\\
Ukrops&Speech&Contact&Overview\\
Enemy&Agency&Intervention&Politician\\
Fire&Path&Expense&Perspective\\
Territory&Psyop&Superiority&Russiahoax\\
Putin&Prigozhin&Exile&Counteroffensive\\
Patriot&Wagner&Journal&Pfizer\\

\bottomrule
\end{tabular}

\end{table}


In a second, more detailed observation, the examination of terms with the highest \textit{keyness} provides insights into the community's perspective, indicating a narrative closely aligned with Russian discourse over time.

On June 23, the emerging narratives, compared to those on June 22, primarily focused on battles in Ukraine, utilizing terms like \textit{battlefield}, \textit{strike} or \textit{territory}. Additionally, posts referred to the Ukrainian army as \textit{ukrops} or \textit{ennemy}, reflecting the Russian argumentation. Moreover, posts portrayed Prigozhin as a \textit{patriot} compared to \textit{putin}.

Moving to June 24, the terms with the highest \textit{keyness} highlighted the mutiny of the \textit{wagner} militia and Putin's \textit{speech}.

On June 25, the key terms underscored Prigozhin's \textit{exile} and Putin's unwavering stance despite the \textit{incident}.

Lastly, on June 26, users discussed the \textit{weakness} of the Russian state, as expressed by various \textit{politician}s from the Republican Party. Additionally, notable terms that reappeared were \textit{pfizer} and the term \textit{russiahoax}, suggesting their perception of a smear campaign against the former US president, due to the decrease in news from Russia.

It is also interesting to note that throughout this period, we find the markers of the supposed cost of American aid to Ukraine with the word \textit{tax} as well as the various \textit{expense}s. In addition,  posts refer to the supposed failure of the Ukrainian \textit{counteroffensive} in line with the Russian state's arguments, as stated by Putin, who claimed that the Ukrainian counter-offensive had failed and that his army had suffered heavy losses\footnote{https://www.bbc.com/news/world-europe-65899424}.

\subsection{Telegram case study}

We proceeded in the same manner as with GAB to detect narratives in Telegram posts and detect the difference between Ukrainian and Russian-oriented news. Similarly, with the narratives found in the Gab analysis, a first observation that can be made on the basis of our \textit{keyness} analysis, is that we can indeed see emerging traces of narratives concerning the attempted mutiny and, as expected, that theses sources focused much more on the war in Ukraine and its consequences. Tables \ref{tab:narrativestableukraine} and \ref{tab:narrativestablerussia} show respectively the top 10 nouns and verbs translated into English with highest \textit{keyness} scores for Ukrainian and Russian Telegram's posts in Russian for the last four days of the dataset (23 June to 26 June).

\begin{table}[htbp]
\centering
\caption{Top 10 nouns and verbs translated into English
with highest \textit{keyness} scores for Ukrainian Telegram's posts in Russian for the last four days of the dataset (23 June to 26 June).}
  \label{tab:narrativestableukraine}
\begin{tabular}{p{16.5mm}p{16.5mm}p{16.5mm}p{16.5mm}}
\toprule
2023-06-23& 2023-06-24&2023-06-25&2023-06-26\\ 
  \midrule
Moscow&Loukachenko&Teenager&Regional council\\
Prigozhin&Wagnerian&NSDC&Construction\\
Wagner&Voronej&Government&Business\\
Channel&Dnipro&Medvedchuk&Performance\\
PMC&Kalibr&Church&Warriors\\
Soldier&Kharkiv&Lugansk&Assets\\
Fighter&Map&Resignation&Wish\\
Group&VKS&Lithuania&Error\\
Employee&Minsk&Shore&Company\\
MO &Guarantee&Dictator&Ukroboron- prom\\
\bottomrule
\end{tabular}

\end{table}

\begin{table}[htbp]
\centering
\caption{Top 10 nouns and verbs translated into English
with highest \textit{keyness} scores for Russian Telegram's posts in Russian for the last four days of the dataset (23 June to 26 June).}
  \label{tab:narrativestablerussia}
\begin{tabular}{p{16.5mm}p{16.5mm}p{16.5mm}p{16.5mm}}
\toprule
2023-06-23& 2023-06-24&2023-06-25&2023-06-26\\ 
  \midrule
Sanction&Rostov&Price&Prison\\
EU&Prigozhin&Recovery&Broadcast\\
Group&Don&Hymn&Ministry of Finance\\
Fact&Headquarter&Apartment&Airplane\\
Regime&Highway&Mediation&Sukhoi\\
Attempt&Support&Dawn&Intelligence\\
Adviser&Building&Epicentre&Analyst\\
Alarm&Moscow Region&Secretary of state&Trump\\
Connection&Situation&Damage&Tax\\
FSB&Road&Militia&Silovik\\
\bottomrule
\end{tabular}

\end{table}

In a second, more detailed observation, the examination of terms with the highest \textit{keyness} provides inside into the difference of narratives spread on Ukraine and Russian Telegram.

As shown in Table \ref{tab:narrativestableukraine}, as early as 23 June 2023, the Ukrainian Channels started to write news about the mutiny attempt as highlighted by words \textit{Prigozhin} and \textit{PMC}. Then, the next day, they continued to talk about the advance towards Moscow of Wagner's troops, noting the town of \textit{Voronej} and the losses suffered by the Russian Air Force (\textit{VKS}), but above all about the aftermath of this attempted mutiny, talking about the agreements reached by \textit{Loukachenko}. Afterwards, on 25 June 2023, news focuses on the consequences of the attempted mutiny on Putin (\textit{Dictator}), such as the supposed \textit{resignation}s of the Chief of Staff and the Russian Defence Minister, and Russia's desire, because of the war, to establish \textit{Medvedchuk} at the head of Ukraine. On 26 June 2023, news focused on Ukrainian companies \textit{Ukroboronprom} and President Zelensky's speech in the Donbas region near the front, \textit{wish}ing the best for his \textit{warriors}. 

For Russian channels, as shown in Table \ref{tab:narrativestablerussia}, we can observe a certain latency in the evocation of events linked to the mutiny. Indeed, on 23 June 2023, they focused on the \textit{sanction}s imposed by the \textit{EU} and the various strikes (\textit{alarm}) against the Kyiv \textit{regime}. On the contrary, on 24 June 2023, all words are connected with the attempted mutiny, updating on the \textit{situation} with places, as \textit{Rostov}-on-\textit{Don} and \textit{Moscow}, and the path, \textit{Highway} and \textit{Road}, taken by \textit{Prigozhin}'s men to reach \textit{Moscow}. For the next two days, the Russian channels focus on restoring order through the values of the state (\textit{Hymn}) and the \textit{Broadcast} of Putin's address to the Russian nation. The news also mentions the \textit{mediation} led by Belarus to stop the mutiny and the damage caused to the \textit{airplane}s of the Russian air force. Finally, it is interesting to note the return of a popular subject in the Russian state media, namely the case of the former US president, \textit{Trump}.

\section{Discussion}

Overall we establish that there is a  significant value in analyzing the time series of \textit{niche} social media and an established messaging service for key narratives.

In line with our hypotheses, the analysis conducted on GAB reveals the presence of narratives related to the Wagner mutiny attempt within communities previously focused on the war in Ukraine and US domestic issues such as Covid vaccination and politics. Also, our inductive approach highlights two key dynamics shaping this transition.

Principally, with our analysis, we observe the flow and ebb of new narratives alongside persistent underlying themes common to this channel. 
These flow and ebb aspects appear to correspond to key global events. In addition, our case study demonstrates the recurrence of certain narratives over time for GAB where peaks in communication occur which include similar narrative elements.  This type of analysis is inherently complex given the challenges of extracting information from big data sources and dynamical multi-party inputs.

Given the unique nature of GAB, these observations provide deeper insights into the characteristics of disinformation narratives. It becomes evident that sustaining disinformation requires a foundation of familiar and recurring elements, while also allowing for adaptation to major global events. On GAB, this ongoing process of recurrence and adaptation is facilitated by the platform's permissive environment, under the banner of freedom of expression.

In a comparative context, our choice of Telegram channels has illuminated discernible narratives. Across different countries, particularly in our selection of news channels for Ukraine and Russia, fresh narratives have emerged in correlation with recent events. However, an important observation pertains to the time lag in reporting the Wagner Mutiny among the Russian Channels compared to the Ukraine channels. Additionally, it's noteworthy that the news posted on Gab appears to be synchronized with the rhythm of Russian developments and propaganda. As an example, on 23 June 2023, both Gab and Russian Telegram channels seem to focus only on \textit{strike}s (\textit{alarm}) which have been carried out in Ukraine and do not mention the Wagner events. Instead, only on 24 June 2023, these two started to include the mutiny and on 25 June 2023, to position the discourse to emphasize the durability of the Russian state despite the \textit{incident}. Finally, on 26 June 2023, both are writing about the US politic in connection with \textit{Trump} and the \textit{Russianhoax}. 


\section{Conclusions and Future Work}

We provide data in support of our mode of time series analysis of social network flows which allows us to extract useful narrative signatures which correlate to events and which develop with time (flow and ebb).  These signatures can be compared across the various channels for a better understanding of the dynamics of narrative detection and development.  We demonstrate the ability to detect evolving narratives on social media with useful characteristics.  This paper offers several contributions. Firstly, it presents a technical pipeline utilizing \textit{keyness} analysis (Log Ratio) to identify traces of narrative innovations and continuity. Secondly, it applies this methodology to the case study of narrative evolution on GAB, a relatively unexplored social network and a known messaging service, Telegram.  Thirdly, given the growth and proliferation of these niche channels like GAB, it's necessary to study them to at least baseline the features. We believe this is the first semi-quantitative and qualitative analysis of GAB.  Numerous prior studies have focused primarily on Twitter \cite{10.1145/3583067} and to some extent YouTube, Telegram \cite{willaert2022detecting} and Reddit. The same approach we present can be done for other niche channels such as GETTR\footnote{https://gettr.com/}, Bitchute\footnote{https://www.bitchute.com/} and Rumble\footnote{https://rumble.com/}. We submit that it is crucial for the broader research community to now incorporate these other channels in the analysis.  This paper can act as an introduction to these channels as very few articles have focused on these new social networks \cite{10.1177/01634437221111943}.    
Finally, and importantly, we suggest that studying the more homogeneous niche social networks can provide elements which can help extract information from the substantially bigger channels. Our initial work evaluating Twitter posts during this period indeed found a detectable Wagner narrative presence but at a substantially reduced overall percentage.  The use of niche channels is a potential mode to extract more signal from the noise in regard to developing methods for social media analysis.  As an outward-looking point, these fringe social media channels can also provide a basis for the behavioral science study of social trends. The key insights provide a basis for a deeper understanding.  It should be emphasized that this exploratory study is limited in scope and our follow-on analysis will look at two key expansion aspects: longer term, non-delta-function events; and also, larger-scale social media time series data.  We anticipate that given the methodology, the general principles will transfer, it is likely that the signal-to-noise may drop, but that will be investigated. 

The application of \textit{keyness} analysis on GAB and Telegram data demonstrates its effectiveness in inductively detecting emerging or persistent narratives, providing valuable insights for further investigation. We acknowledge the limitations of the exploratory scope of this paper and suggest the results as a starting point for further research.

To enhance the analysis, future research could incorporate wider semantic networks by employing statistically-informed co-occurrence analyses. Additionally, introducing more granularity through graph-like representations of narratives inferred from argument structures could offer promising avenues for contextualizing key items. 

On a conceptual level, this analysis prompts broader questions of meaning and interpretation. The \textit{keyness} analysis itself does not capture the semantic nuances of the examined posts. Therefore, assigning meaning to key items requires human interpretation, such as considering combinations of key items, referring to the corpus for keyword exploration, and contextualizing within cultural and media-theoretical frameworks. 

The main contributions of this paper are to provide a type of methodology for \textit{deciphering} time series events in social media. These initial results do not allow for broad generalizations, but are indicative.  The potential limitations can also be noted and potentially addressed.  For example, the keyness analysis, while quantitatively robust, might miss some of the (human perceived) deeper semantic meanings and nuances that develop in online narratives.  These limitations highlight the extreme challenges to narrative time series analysis.  However, a simplified case study such as presented, allows for a mode of attack while determining the extent of the applicability.  

This raises the need for the development of critical frameworks that integrate inductive methods. Proposals such as \textit{data hermeneutics} have been introduced \cite{10.1007/s00146-018-0856-2}, providing opportunities for future research to explore actionable implementations in this domain. Moreover, it will be interesting to analyse images and videos contained in posts to help the contextualization \cite{li2023blip2}.

\bibliography{aaai24}

\end{document}